\documentclass[twocolumn,showpacs,preprintnumbers,amsmath,amssymb]{revtex4}
\usepackage{graphicx}
\usepackage{dcolumn}
\usepackage{bm}
\begin{document}

\title{Asymmetric localization in disordered Landau bands}
\author{A. Aldea and M. Ni\c{t}\u{a}}
\address{Institute of Physics and Technology of Materials, POBox 
MG7, Bucharest-Magurele, Romania}
\date{\today}
\begin{abstract}
We show that
due to the Landau band mixing the eigenstate localization
 within the disordered bands get
an asymmetric structure: 
the degree of localization increases in the
lower part of the band and decreases in the upper one.
The calculation is performed  for a 2D lattice with the Anderson disorder
potential and we prove that this effect is related to the upper shift of
the extended states within the band 
and is enhanced by the disorder strength.
The asymmetric localization and the energy shift dissapear when the interband
coupling is switched off.

 

\end{abstract}
\maketitle


The localization effect in Landau bands
attracted much interest since the discovery of 
the QHE \cite{halperin1982, aoki1982, ono1982, ando1983, levine,
aoki1985:831, chalker, huckestein, huo}.
When the impurity potential is present,
the initially degenerate Landau levels of the 2D system 
are turning into broad bands.
In contrast  to the zero magnetic field case when there are no extended states 
in two dimensions \cite{abrahams1979},  extended states are present
at the center of each Landau energy band \cite{halperin1982}.
The early studies performed for the continuous Hamiltonian model
with neglected inter-Landau band mixing 
\cite{aoki1982, ono1982, ando1983}
show that the generic picture of the 2D Landau bands contains
 localized states in  the band tails and extended states in the middle.
However, in a real 2D system the band separation may become smaller
than the band width so that the above approximation is not always valid.
It has been shown that the mixing between different Landau bands, that 
generally comes from the disorder presence,
is related to the energy shift of the extended states from
the central position of the band
\cite{ando1989, liu, shahbazyan, kagalovsky, gramada, haldane, koschny2003}
or it may have a delocalization effect when the states with opposite
chirality are coupled \cite{xiong}.

The lattice model  captures  this relevant feature, and this happens
because the discrete Landau model (initially solved by Hofstadter 
for the pure case \cite{hofstadter}) takes automatically into account
the inter-band coupling. The energy shift between the position of the
extended states and the peak of the density of states was explicitely
calculated in the lattice model by \cite{pereira}.
Furthermore, for the same model,
one notes the asymmetric behavior of the 
eigenstates localization in Landau bands. 
This result was reported in \cite{aoki1985} by the calculation
of the localization length
within the outer Landau band,
and is consistent with the asymmetric behavior of the inverse 
participation number within the Landau bands reported recently in \cite{noipss}.
In this paper we study the origin of this asymmetry.
By the projection of the 2D discrete Hamiltonian on the nondisordered Landau levels we
show here that this asymmetric behavior of the eigenstates localization
are due to the interband coupling terms. It will be also shown that
this effect
is intimately related to the shift of the extended states. It means that 
both of the two effects disappear  when the interband coupling is switched off.

We use for this
the following 2D Landau Hamiltonian written in the
discrete basis of a 2D rectangular lattice:
\begin{eqnarray}\label{h}
H(\phi)=&&\sum_{n=1}^{N}\sum_{m=1}^{M}
t~e^{2\pi i m \phi}\vert n,m\rangle\langle n+1,m\vert \nonumber\\
&&+t\vert n,m\rangle\langle n,m+1\vert + H.c.~\big] +V,
\end{eqnarray}
where V is the Anderson disorder potential:
\begin{eqnarray}\label{v}
V=\sum_{n=1}^{N}\sum_{m=1}^{M}{\epsilon}_{nm} \vert n,m
\rangle\langle n,m\vert.
\end{eqnarray}

The discrete points ($n,m$) define the 2D rectangular lattice with surface
$L^2=N\times M$ and lattice constant $a$. 
\{$|n,m\rangle$\} with $n=1\cdots N$ and $m=1\cdots M$ is the discrete 
vector basis
and generates the Hilbert space of the one-electron states. 
The periodical boundary 
conditions are used, meaning that $|n,M+1\rangle=|n,1\rangle$ and
$|N+1,m\rangle=|1,m\rangle$ (2D toroidal geometry).
The perpendicular magnetic field in the Landau gauge $A=(-By,0,0)$
 is introduced by the Peierls substitution in the hopping elements 
 along the $x=n a$ direction, $t\to t \exp\{2\pi i m \phi\}$,
where $\phi$ is the magnetic flux through the unit cell 
$a^2$ of the lattice measured in the quantum flux units $\phi_0=h/e$.
In (\ref{v}) the energies $\epsilon_{nm}$ represent the random variables 
uniformly distributed in the energy interval $[-W,W]$. $W$ is the amplitude of 
the Anderson disorder potential (or disorder strength); t is the energy unit and is set to 1.
For the commensurate values of the magnetic flux, the eigenstate spectrum 
of the pure system (V=0)
exhibits the well-known Hofstadter butterfly structure \cite{hofstadter}.
In the numerical calculation we set the flux value  as the ratio
 $\phi=1/p$ and the system size as $L^2=(integer\cdot p)^2$. 
In this case the eigenstates of the 2D system 
 are grouped in $p$ discrete Landau bands,
every band having a number of $n_b=L^2/p$ degenerated eigenstates.
To each energy level $\epsilon_{\alpha}^0$, with the band index $\alpha=1
\cdots p$, correspond $n_b$ degenerate eigenvectors 
$|\Psi_{\alpha i}^0\rangle=|\alpha i\rangle$ with $i=1\cdots n_b$.
As the system has the electron-hole symmetry
we concentrate on the lower half of the spectrum.

In the presence of the disorder potential the degenerate energy level 
$\epsilon_{\alpha}^0$ turns into the broad energy band 
$\{\epsilon_{\alpha i}\}$ with $i=1\cdots n_b$. 
 We study the degree of localization  of the
nondegenarate eigenstates $|\Psi_{\alpha i}\rangle$
 by the calculation of the inverse participation numbers IPN which is 
defined as:
\begin{eqnarray}\label{ipn}
IPN= P_{\alpha i}=\sum_{n,m}|\langle n,m|\psi_{\alpha i\rangle}|^4.
\end{eqnarray}
$P_{\alpha i}$ varies from $1/L^2$
for the extended states, when the electron wave function spreads out 
over the whole
surface of the plaquette to $1$ for the strong localized states.
The nature of the eigenstates  can also be checked  by the calculation of
the variance of the level spacing distribution, $\delta t$ \cite{noiprb}. 

In this work 
we put into evidence the role of the interband interaction.
To this  end we write the Hamiltonian (\ref{h})
in the vector basis of the 2D pure system, $\{|{\alpha i\rangle}\}$:
\begin{eqnarray}\label{hb}
H(\phi)=&&\sum_{\alpha=1}^{p}\sum_{i=1}^{n_b}
{\epsilon}_{\alpha}^0 
\vert \alpha i\rangle\langle \alpha i\vert+
\sum_{\alpha=1}^{p}\sum_{i,j=1}^{n_b}
V_{\alpha i,\alpha j} \vert \alpha i\rangle\langle \alpha j\vert
\nonumber\\
&&+c\sum_{\alpha \ne \beta=1}^{p}\sum_{i,j=1}^{n_b}
V_{\alpha i,\beta j} \vert \alpha i\rangle\langle \beta j\vert.
\end{eqnarray}
In this representation the disorder potential $V$ becomes a sum
of two terms, corresponding to the inter- and intra-band coupling
(the second and the third term in (\ref{hb}), respectively).
$V_{\alpha i,\beta j}$ are
the matrix elements of the Anderson potential $V$ written in the basis
of the eigenfunctions of the ordered system $\{| \alpha i\rangle \}$. 
They are random variables as well, and their values are proportional to
the disorder strength $W$. The coupling constant $c$ is introduced 
for convenience.

\begin{figure}
\includegraphics[scale=0.6]{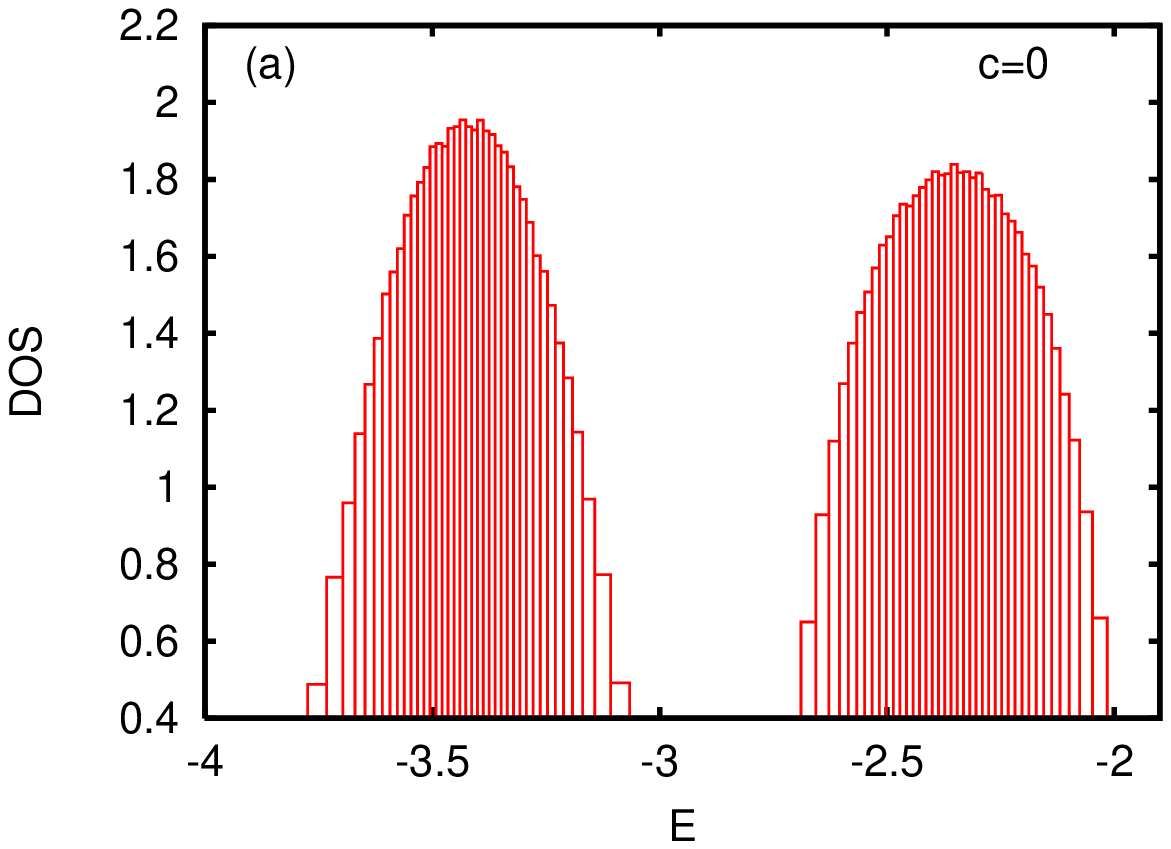}
\includegraphics[scale=0.6]{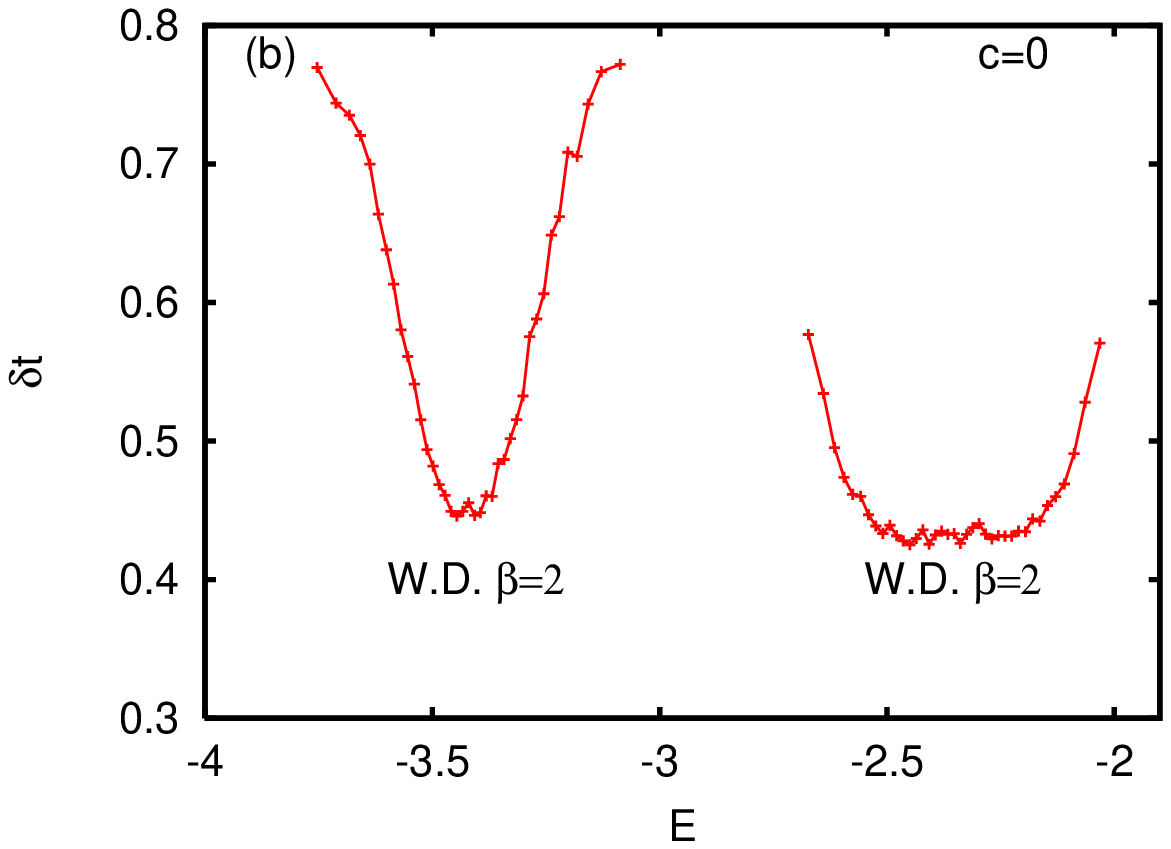}
\includegraphics[scale=0.6]{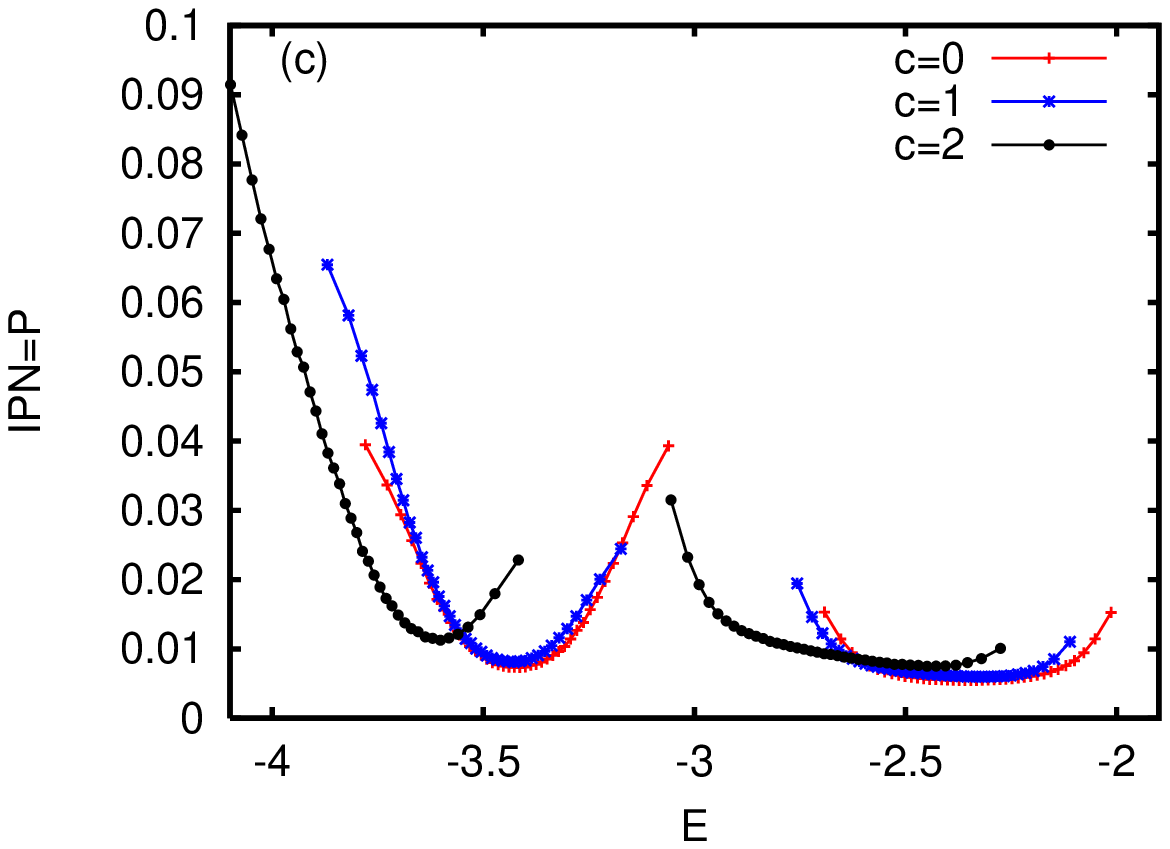}
\caption{The density of states (a), the level spacing variance
 $\delta t$ (b), and the inverse participation number (c)
 for the first two Landau bands vs. energy.
($L^2=20^2$, $W=2$, $\phi$=0.1). DOS and $\delta t$ are presented 
only for the symmetric case with interband coupling constant c=0. 
IPN is presented for c= 0, 1 and 2. 
In (b), when $\delta t=\sqrt{\frac{3\pi}{8}-1}\simeq 0.42$, 
the eigenstates correspond to the extended states of the
unitary Wigner Dyson ensemble ($\beta=2$). 
(The configuration average is performed over 5000 samples).
}
\end{figure}

We shall discuss first the situation when the parameter c is set to zero in (4)
, meaning that only the intraband coupling is taken into account.
A picture of the disordered bands for this case
is given in Fig.\,1. The density of states (DOS) for the first two bands 
is shown in Fig.\,1a and its profile has a semielliptic shape.
Let $E_b$ be the energy where DOS reaches its maximum, which in this case 
is located in the middle of the band.
The level spacing distribution is calculated by averaging
over different disorder configurations in the manner described in \cite{noiprb}.
It is known that the extended states belong to the unitary Wigner-Dyson 
ensemble $\beta=2$ with the variance of the level spacing $\delta t=0.42$,
while the localized ones are distributed according to the Poisson law 
 with the variance  $\delta t=1$ \cite{mehta}.
The calculation of the level spacing variance in Fig.\,1b shows the presence 
of extended eigenstates  in the middle of the disordered bands, 
for energies $E$ around $E_b$.
For energies towards the band edges the states become localized and
there is a continuous crossover from unitary Wigner-Dyson distribution to the
Poisson distribution as in \cite{noiprb}. 
Obviously, the most localized states are at the edges of the  band, 
where $\delta t\ $ increases to higher values.
We note  that in the thermodynamic limit the extended states in the 
band center collapse into a single energy level \cite{huo}.

\begin{figure}
\includegraphics[scale=0.6]{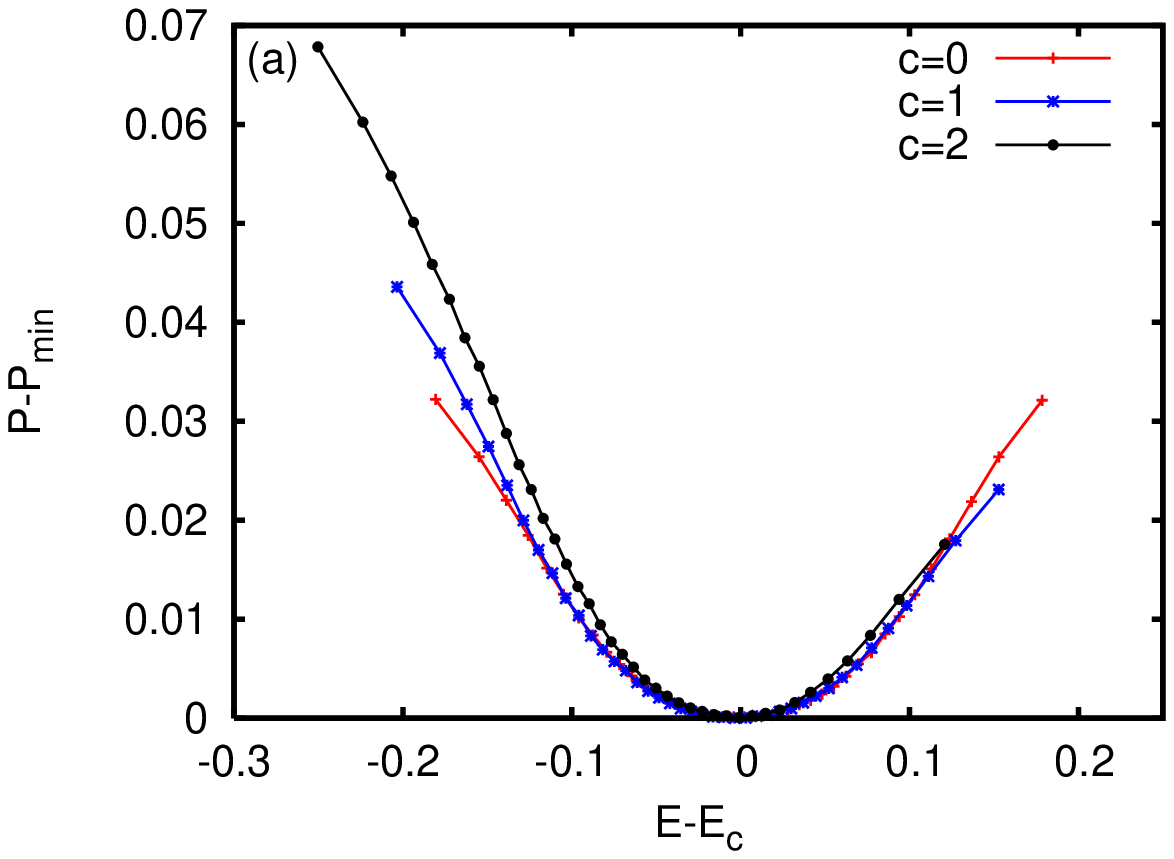}
\includegraphics[scale=0.6]{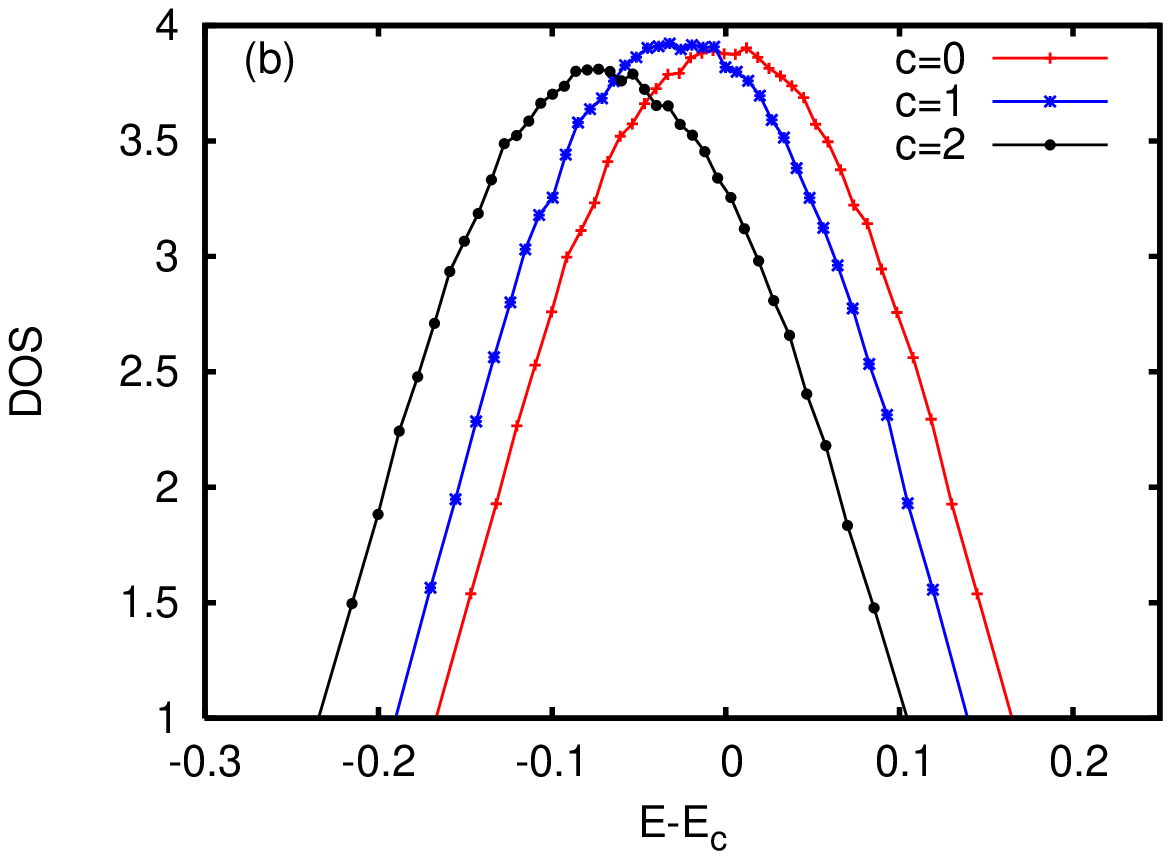}
\caption{(a) The inverse participation number $P-P_{min}$ and (b) 
the density of states
for the first Landau band  plotted vs. energy  at different values of the
interband coupling constant: c=0 (crosses), c=1 (stars) and c=2 (dots).
($L^2=20^2$, $W=1$, $\phi=0.1$).
$P_{min}$ is the minimum value of the inverse participation number, while 
$E_c$ is the energy where this value is reached, i.e. $P_{min}=P(E_c)$.}
\end{figure}

We complete the picture of the eigenstates localization for uncoupled bands
($c=0$) showing in Fig.\,1c (the red curve) the values of the  IPN. 
By varying the energy from the central position of every band, the 
IPN  increases, indicating an increased localization of the eigenstates. 
Let $E_c$ be the energy with the lowest IPN value in the middle of the band.
 This is the energy of the most extended state, where, in the
thermodynamic limit, the localization-delocalization transition takes place.
For the case discussed here, when c=0,
the critical energy corresponds to the maximum of the DOS ($E_c=E_b$), and
the IPN is symmetric within the band. This is what we call the symmetric case.
These properties are  not preserved  anymore when $c\ne 0$ as we could  already 
notice in Fig.\,1c (see the IPN curves for c=1 and 2). 

In what follows, we are interested in finding out how the localization 
properties evolve with the interband coupling c.
In Fig.\,2 we show the result of the numerical calculation 
for the inverse participation number and  density of states  
as function of $E-E_c$ for three values of the interband coupling 
constant c= 0, 1 and 2.  
For the first band, the IPN curves 
are depicted in Fig.\,2a. By the definition of $E_c$, the IPN
takes the minimum value at $E-E_c=0$. 
One remarks  that the symmetry of the IPN  
 is lost for nonvanishing coupling constant  c=1 and 2, 
i.e. in the case of the band mixing.
Compared to the uncoupled case (c=0), the increased values of IPN 
for $E<E_c$ at $c\ne 0$
indicate an increased degree of localization in the lower part of the band.
The opposite is true in the upper part where the states become less localized.

In Fig.\,2b we depict the DOS of the first Landau band. For c=1, the 
band  is shifted downwards meaning that the maximum of DOS 
does not correspond to  $E_c$, but it occurs at a lower energy $E_b < E_c$.
The shift increases for c=2. It means that the critical energy 
is moving up in the band when the interband coupling  is present. 
This asymmetric behavior is preserved for all the bands 
contained in the lower half of the spectrum.  

Once we have established the relation between the interband coupling and the 
asymmetry of the Landau bands, we are interested now to learn how this property 
depends on disorder.
We  keep fixed c=1 in (\ref{hb}) and increase the amplitude of disorder $W$.
Since the interband coupling in the discrete Hamiltonian is due to the presence
 of the disorder, we expect the shift of the critical energy be also 
dependent on the disorder amplitude.

Fig.\,3 gives the inverse participation number and the 
density of states as function of $(E-E_c)/W$ for different values of $W$. 
One notices in Fig.\,3a  that, for any disorder, in the domain of extended 
states around $E_c$ the inverse participation number $P(E)$ can be expressed as 
$P(E)-P(E_c)=f((E-E_c)/W)$.
Deviations from this law are noticed at the band edges. 
By increasing the disorder amplitude $W$, the lower energy states of the 
band become more localized, but the higher energy states of the band 
become less localized.
At the same time, the extended states are moving towards the upper edge 
of the band, so that the critical energy $E_c$ does not  
correspond any more to the maximum of the density of states.  
This can be seen in Fig.\,3b,  where one notices the band shift with  
increasing  disorder strength.
The critical energy shift $E_c-E_b$ as function of disorder
is shown in Fig.\,4 for the first two bands in the spectrum.
For the present model, $(E_c-E_b)/W$ exhibits a linear dependence on the 
disorder strength $W$. We note no significant difference 
of the critical energy shift for the two bands depicted.  
The numerical calculations  have been repeated for an increased system size, 
the results being the same.

\begin{figure}
\includegraphics[scale=0.6]{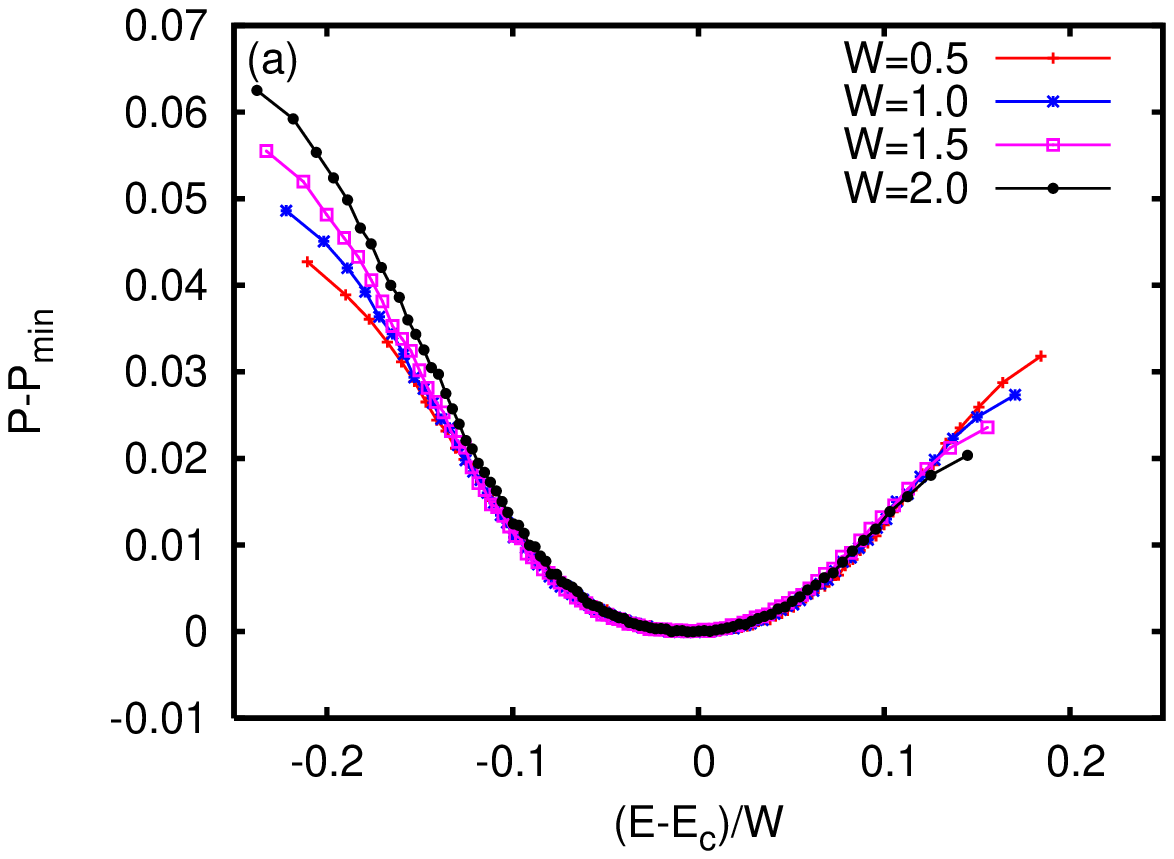}
\includegraphics[scale=0.6]{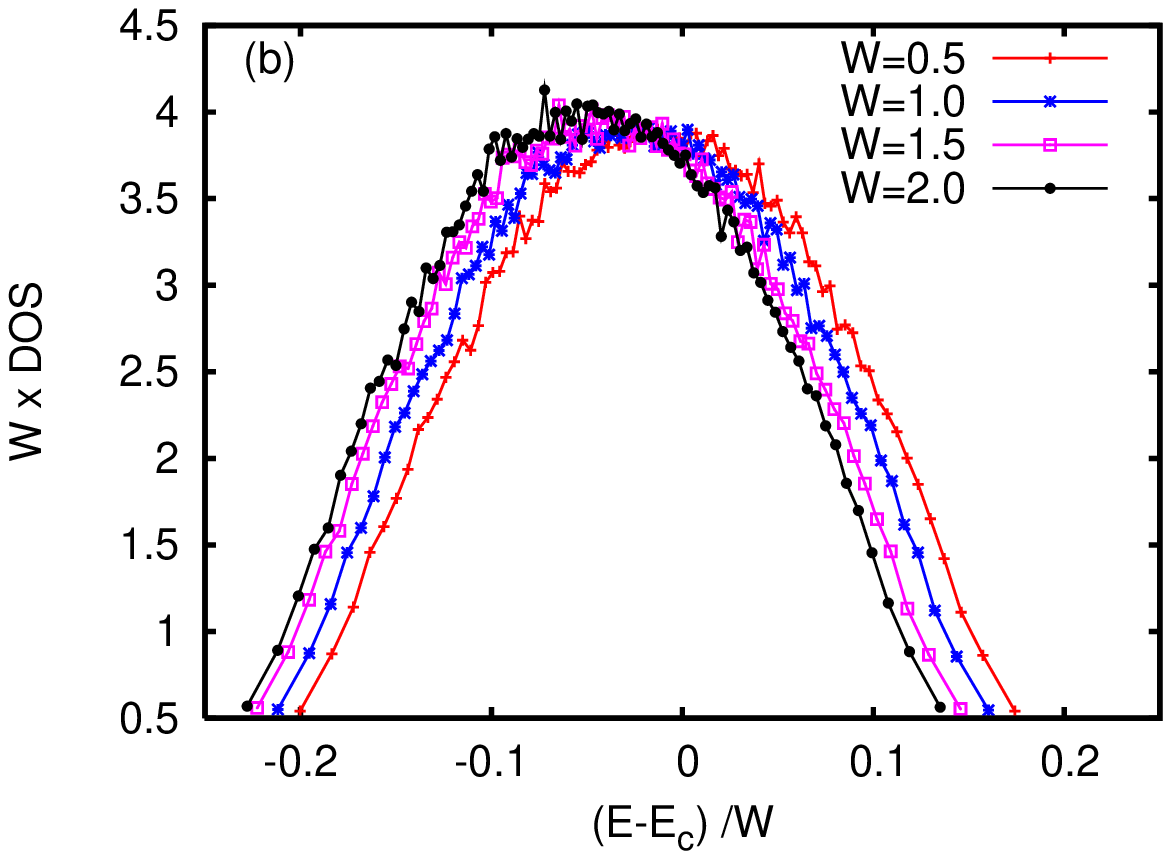}
\caption{
(a) The inverse participation number $P-P_{min}$, and (b) the density of states 
DOS for the first Landau band vs. energy at different disorder strength
 W=0.5 (crosses), W=1 (stars), W=1.5 (squares), and W=2.0 (dots).
($L^2=30^2, \phi=0.1, c=1$).
$P_{min}$ is the minimum value of the inverse participation number, while
$E_c$ is the energy where this value is reached, i.e. $P_{min}=P(E_c)$.
Note that the energies are scaled by the disorder strength $W$.
}
\end{figure}

\begin{figure}
\vspace{0.3 cm}
\includegraphics[scale=0.6]{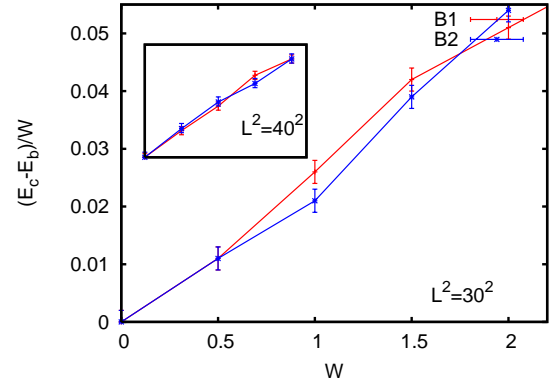}
\caption{The shift of the critical energy $E_c$ vs. disorder strength W.
($L^2=30^2$, $\phi=0.1$, c=1).
The results are plotted for the first band (B1) and for the second band (B2).
In the inset are represented the numerical results obtained for $L^2=40^2$.
}
\end{figure}

In conclusion, we have shown that the band mixing gives rise to the 
asymmetry  of the localization properties in the Landau bands.
The inverse participation number
(which measures the degree of localization) becomes an asymmetric function
within the band, indicating that the degree of localization increases
for the states in the lower energy part of the band and decreases
for the states in the upper  part.
At the same time the critical energy (the most extended state) in each Landau band
does not correspond to
 the maximum of the density of states but it is shifted  to higher energies.
These properties are specific to the many-band model 
and are quite different from  the properties of the one-band model which 
exhibits only symmetrical features. In a large range of energies
the inverse participation number $P(E)-P(E_c)$ can be expressed as a universal
function of $(E-E_c)/W$. 
%
%
%


The shift of the extended states from the central position of the Landau band
supports the result reported recently by Shlimak et al \cite{shlimak} consisting
in the displacement of the integer values of the filling factor 
from the middle point of the QH plateaus in disordered Si-MOSFET samples.

\begin{acknowledgments}
We are grateful to I.Shlimak for drawing our attention on this topic. 
This work was supported by the National Programme for Basic Research
and  Sonderforschungsbereich 608.
\end{acknowledgments}

\end{document}